\documentclass[preprintnumbers,superscriptaddress,amsmath,amssymb,prd,nofootinbib,onecolumn]{revtex4-2}
\usepackage{amssymb}
\usepackage{mathrsfs}
\usepackage{amsthm}
\usepackage{braket}
\usepackage{bm}
\usepackage{qcircuit}
\usepackage{graphicx}
\usepackage{amsfonts}
\usepackage[colorlinks,linkcolor=red,anchorcolor=blue,citecolor=green]{hyperref}
\usepackage{subfigure}
\usepackage{float}
\usepackage{enumerate}
\usepackage{multirow}

\usepackage{array}
\usepackage[figuresright]{rotating}
\usepackage{upgreek}

\begin{document}
	
	\title{Acceleration of the Universe without the Hubble tension with \\ Kaniadakis holographic dark energy using the Hubble horizon as the IR cut-off}
	
	\author{Wei Fang}
	\email{wfang@shnu.edu.cn}
	\affiliation{Department of Physics, Shanghai Normal University,
		100 Guilin Rd, Shanghai 200234, P.R.China}
	\affiliation{The Shanghai Key Lab for Astrophysics, 100 Guilin Rd, Shanghai 200234, P.R.China}
	
	\author{Guo Chen}
	\thanks{Co-first author.}
	\affiliation{Department of Physics, Shanghai Normal University,
		100 Guilin Rd, Shanghai 200234, P.R.China}
	
	\author{Chao-Jun Feng}
	\thanks{Corresponding author}
	\email{fengcj@shnu.edu.cn}
	\affiliation{Department of Physics, Shanghai Normal University,
		100 Guilin Rd, Shanghai 200234, P.R.China}
	
	\author{Wei Du}
	\email{wfang@shnu.edu.cn}
	\affiliation{The Shanghai Key Lab for Astrophysics, 100 Guilin Rd, Shanghai 200234, P.R.China}

	\author{Chenggang Shu}
	\affiliation{The Shanghai Key Lab for Astrophysics, 100 Guilin Rd, Shanghai 200234, P.R.China}

	\begin{abstract}
		We introduce a holographic dark energy model that incorporates the first-order approximate Kaniadaski entropy, utilizing the Hubble horizon, $1/H$, as the infrared cutoff. We investigate the cosmological evolution within this framework. The model introduces an extra parameter relative to the $\Lambda$CDM model. It posits a Universe that is initially dominated by dark matter, which then evolves to a phase where dark energy becomes the predominant component, with this transition occurring at a redshift of approximately $z \sim 0.419$. The energy density of dark energy is ultimately expected to become constant, thereby circumventing the potential issue of a "big rip." Employing the most recent Type Ia supernova and Hubble parameter data, we constrain the model's parameters and find a Hubble constant of $H_0=72.8$ km/s/Mpc, thereby resolving the Hubble tension issue. The estimated age of the Universe, based on the best-fit parameter values, is $14.2$ Gyr. Furthermore, we predict the number of strong gravitational lenses and conduct statefinder and $Om$ diagnostic analyses to validate and characterize the model.
	\end{abstract}

	\maketitle

	\section{Introduction}
	Observations indicate that the Universe is undergoing accelerated expansion, and to explain this phenomenon, many dark energy models have been proposed. Certainly, the simplest dark energy model is the cosmological constant, known as the $\Lambda$CDM model. In this model, the energy density of dark energy is a constant, or equivalently, its equation of state parameter $w = -1$. However, this model suffers from a fine-tuning problem, and recent observations suggest that the equation of state parameter of dark energy is not a constant. Therefore, at present, the $\Lambda$CDM model may not be sufficient to describe the current cosmological accelerated expansion phenomenon, or at least, it is not that simple.
	
	Among the myriad of dark energy models, the holographic dark energy model has emerged as a focal point of interest and academic exploration \cite{LiMiao2004}. This model is underpinned by the holographic principle, which asserts that the entropy of the Universe should not surpass that of a black hole of equivalent size. As a result, the energy density, which is proportional to the inverse square of an infrared cutoff $\sim 1/L^2$, is constrained. This principle has recently been explored in the context of thermodynamics \cite{Manoharan2023}. Furthermore, the Rényi holographic dark energy model, which examines the interaction between dark energy and dark matter in fractal cosmology, is discussed in Ref.\cite{Zhang2022}. In Ref. \cite{Feng:2009jr}, the authors also incorporated viscous fluid, significantly mitigating the issue of the Universe's age. In Ref. \cite{Feng:2008hk}, the authors reconstructed the $f(R)$ theory model from the holographic dark energy, with recent advancements on the holographic dark energy model available in \cite{Wang:2016och} and information on dark energy and modified gravity in Ref.\cite{Bahamonde:2017ize}.
	
	Kaniadakis introduced a single-parameter generalization of the Boltzmann-Gibbs entropy, known as the Kaniadakis entropy, which is defined as:
	\begin{equation}
		S_{k}=-k_{B}\sum_{i}n_{i}\ln_{{k}}{n_{i}}\label{2.1} \,,
		\quad \text{or} \quad S_{k}=-k_{B}\sum_{i}^{W}\frac{P_{i}^{1+K}-P_{i}^{1-K}}{2K} \,,
	\end{equation}
	where \(P_{i}\) represents the probability of a specific microstate of the system, and \(W\) denotes the total number of possible configurations. When applied within the context of black hole physics, the entropy takes the form
	\begin{equation}
		S_{k}=\frac{1}{K}\sinh (KS_{BH}) \label{2.3} \,,
	\end{equation}
	which, in the approximation where \(K\ll 1\), can be simplified to
	\begin{equation}
		S_{K}=S_{BH}+\frac{K^{2}}{6}S_{BH}^{3}+ \mathcal{O}(K^4)\label{eq:entropy}\,.
	\end{equation}
	Drawing from the holographic principle, we can derive the so-called Kaniadakis holographic dark energy (KHDE), whose energy density is expressed as
	\begin{eqnarray}
		\rho_{de} = 3(\alpha L^{-2} +\tilde{\beta}  L^2)\,,
	\end{eqnarray}
	where \(L\) serves as an infrared cutoff. In literature, the future event horizon has frequently been selected as the cut-off boundary, as seen in \cite{Drepanou2022}. Specifically, for the KHDE model, \(L = R_h = a\int_t^\infty ds/a(s)\) is utilized, a method also referenced in \cite{Ghaffari2022} within the Brans-Dicke framework. Furthermore, the dynamic characteristics of Kaniadakis holographic dark energy were scrutinized in \cite{Hernández2022}. In another study \cite{Luciano2022}, the authors delved into the applicability of Kaniadakis statistics as a leading paradigm for characterizing intricate systems within the realm of relativity. The justification behind opting for the future event horizon is rooted in the fact that the Hubble horizon, in the initial holographic dark energy model, is inadequate in propelling the Universe's accelerated expansion. Nevertheless, the incorporation of an additional term in Equ.~\eqref{eq:entropy} modifies this scenario.

	In this paper, we utilize the Hubble horizon as the infrared cutoff for the Kaniadakis holographic dark energy, which is derived from the first-order approximation of the dark energy density derived from Kaniadakis entropy. We examine the evolutionary trajectory of the Universe and find that our model successfully accounts for the observed accelerated expansion. After performing the parameter fitting procedure, we also discover that the current Hubble parameter $H_0=72.8$ km/s/Mpc, hence there is no Hubble tension in this model.  For recent works on the Hubble tension, please refer to \cite{Liu:2024vlt}, \cite{froggatt2024domain}, and \cite{He:2024jku}. The transition from matter dominance to dark energy dominance approximately occurs at a redshift of $z \approx 0.419$.
	The estimated age of the Universe, based on the best-fit parameter values, is $14.2$ Gyr. Furthermore, we  predict the number of strong gravitational lenses and conduct statefinder and $Om$ diagnostic analyses to validate and characterize the model. In the future, the energy density of dark energy will asymptotically approach a constant, behaving much like the cosmological constant, and hence, there will be no issue of a cosmic "big rip."
	
	In the following section, we will delve into the cosmological evolution of the holographic dark energy model and subsequently constrain the model's parameters using empirical data. Moreover, we have derived exact analytical solutions to the Friedmann equations, which we have employed to estimate the age of the Universe. Subsequent to this, we  predict the number of strong gravitational lenses and examine the statefinder and Om diagnostic parameters. In the final section, we present our conclusions  of the findings.

	\section{Kaniadakis Holographic Dark Energy with the Hubble Horizon as the IR Cut-off}
	
	In the following, we will consider a flat, homogeneous, and isotropic  Universe described by the Friedmann-Robertson-Walker metric
	\begin{eqnarray}
		ds^2 = -dt^2 + a^2(t)(dr^2+r^2d\theta^2 + r^2\sin^2\theta d\phi ^2) \,.
	\end{eqnarray}
	After taking the  Hubble horizon as the IR cut-off, $L = 1/H$, the energy density of the Kaniadakis holographic dark energy is
	\begin{eqnarray}\label{eq:de}
		\rho_{de}=  3\alpha H^2 +3\tilde{\beta}  H^{-2} \,,
	\end{eqnarray}	
	where $\alpha$ is a dimensionless constant, $\tilde{\beta}$ has the dimension of mass$^4$  and the factor of 3 introduced before the two constants is for convenience. Then the Friedmann equation becomes
	\begin{eqnarray}\label{eq:fred1}
		3H^2 = \rho_m + \rho_r +  3\alpha H^2 +3\tilde{\beta} H^{-2} \,,
	\end{eqnarray}	
	where  $\rho_m = \rho_{m0}a^{-3}$ and $\rho_r =\rho_{r0}a^{-4}$ are the energy densities of matter and radiation, respectively, and $H$ is the  Hubble parameter. Here and henceforth, we adopt the unit \(8\pi G = 1\).
	
	In the later stages of the Universe's evolution, radiation can be neglected because it decays much more rapidly than other components. Therefore, in the following, we only consider two components: dark matter and dark energy until we perform the observational constraints.  Slightly rearranging the Friedmann equation, we can obtain
	\begin{eqnarray}\label{eq:fried}
		3(1-\alpha)H^4  -\rho_m H^2 -3\tilde{\beta}  = 0 \,.
	\end{eqnarray}
	When $\alpha = 1$ and $\beta < 0$, the above equation indicates that the Hubble parameter increases as the Universe expands: $H^2= (-3\tilde{\beta} /\rho_{m0} )a^3$. This is inconsistent with observations. So we only consider the solution with $\alpha \neq 1$:
	\begin{eqnarray}\label{eq:sol1}
		H^2 = \frac{\rho_m\pm \sqrt{\rho_m^2 +36\tilde{\beta} (1-\alpha)}}{6(1-\alpha)} \,.
	\end{eqnarray}
	Observations show that the Universe transitions from being matter-dominated to dark energy-dominated. Therefore, the solution above (\ref{eq:sol1}) should take a positive sign
	\begin{eqnarray}\label{eq:sol2}
		H^2 = \frac{\rho_{m0}a^{-3} + \sqrt{\rho_{m0}^2a^{-6} +36\tilde{\beta} (1-\alpha)}}{6(1-\alpha)} \,.
	\end{eqnarray}
	Thus, to ensure that \(H^2 > 0\), \(\alpha\) must be less than 1.  From Equ.~\eqref{eq:sol2} , it can be seen that in the early Universe when $a \ll 1$, $H^2 =  \frac{1}{3}\rho_{m0}a^{-3}/(1-\alpha)$. As $a$ increases, meaning the matter component continuously decays, the Universe will tend towards a dark energy-dominated state, where the energy density of this dark energy becomes a constant, i.e.
	\begin{eqnarray}
		H^2 \rightarrow \sqrt{\tilde{\beta}/(1-\alpha)}\,, \quad a\rightarrow \infty\,.
	\end{eqnarray}
	
	Substituting Equ.~\eqref{eq:sol2} into Equ.\eqref{eq:de}, we obtain
	\begin{eqnarray}
		\rho_{de} 
		\nonumber
		&=& \frac{\alpha }{2(1-\alpha)} \left( \sqrt{\rho_m^2 +36\tilde{\beta} (1-\alpha)}+\rho_m \right)  +  \frac{1} { 2} \left (\sqrt{\rho_m^2 +36\tilde{\beta} (1-\alpha)} - \rho_m \right ) \\
		&=&\frac{1}{2(1-\alpha)} \sqrt{\rho_m^2 +36\tilde{\beta} (1-\alpha)} + \frac{2\alpha - 1}{2(1-\alpha)} \rho_m\,.\label{eq:rho}
	\end{eqnarray}
	From the continuity equation for dark energy, we also get the pressure of the dark energy as follows
	\begin{eqnarray}
		p_{de} = -\frac{\dot \rho_{de}}{3H} - \rho_{de}  
		&=&  \frac{1}{2(1-\alpha)}
		\left[ \frac{\rho_m^2 }{ \sqrt{\rho_m^2 +36\tilde{\beta} (1-\alpha)} }
		-\sqrt{\rho_m^2 +36\tilde{\beta} (1-\alpha)}  \right]\,,
	\end{eqnarray}
	where we have used $\dot \rho_m = -3H\rho_m$.  Then, in the era when dark matter dominated  the energy density of the dark energy evolved asymptotically to $\rho_{de}  \rightarrow \alpha \rho_m /(1-\alpha)$ and it exhibited zero pressure. The equation of state can be obtained $w_{de} = p_{de}/\rho_{de}$.  Using Equ.~\eqref{eq:fried} and taking its derivative, the deceleration parameter of the Universe can be given by the following equation:
	\begin{eqnarray}
		q = -1- \frac{\dot H}{H^2} =-1 + \frac{3}{2} \rho_m \bigg[ \rho_m^2 +36\tilde{\beta} (1-\alpha)  \bigg]^{-1/2} \,.
	\end{eqnarray}
	It is clear that as $\rho_m$ decays to zero, the deceleration parameter approaches $-1$.
	
	\section{Observational constraints}
	Define a dimensionless Hubble parameter, 
	\begin{eqnarray}\label{eq:fred2}
		E^2 = \frac{H^2}{H_0^2} = \frac{\Omega_{mr}(z)+  \sqrt{\Omega_{mr}^2(z) +4\beta(1-\alpha)}}{2(1-\alpha)} \,,
	\end{eqnarray}
	where $\beta = \tilde{\beta}/H_0^4$ is a dimensionless constant and 
	\begin{eqnarray}
		\Omega_{mr} (z)= \Omega_{m0}(1+z)^3 + \Omega_{r0}(1+z)^4 \,,
	\end{eqnarray}
	with $z$ the redshift. And we have also defined  dimensionless parameters
	\begin{eqnarray}
		\Omega_{m0} = \frac{\rho_{m0}}{3H_0^2} \,, \quad 	\Omega_{r0} = \frac{\rho_{r0}}{3H_0^2}\,,
	\end{eqnarray}
	for matter and radiation respectively. 
	From Equ.~\eqref{eq:fred1}, we obtain 
	\begin{eqnarray}\label{eq:beta}
		1= \Omega_{m0}+\Omega_{r0} + \alpha + \beta \,,
	\end{eqnarray}
	which means there are only three free parameters in this model: $\Omega_{m0}, \Omega_{r0}$ and $\alpha$.  In the following, we will perform the constraint on these parameters using observational data sets based on Markov Chain Monte Carlo method.
	
	We employ the Pantheon compilation of Type Ia supernova (SNIa) data \cite{Pan-STARRS1:2017jku} and Hubble parameter ($H(z)$) data points \cite{Sharov:2018yvz} to constrain the model parameters. The SNIa dataset encompasses 279 samples from the Sloan Digital Sky Survey (SDSS) and the Supernova Legacy Survey (SNLS), spanning a redshift range of $0.03 < z < 0.68$. Additionally, it includes 1048 samples with redshifts ranging from $0.01 < z < 2.3$, incorporating the Hubble Space Telescope (HST) samples and various low-$z$ samples. The $H(z)$ data consists of 26 data points derived from Baryon Acoustic Oscillations and 31 data points obtained through the differential age method.
	
\begin{figure}[htbp]
		\centering
		\includegraphics[width=0.7\linewidth]{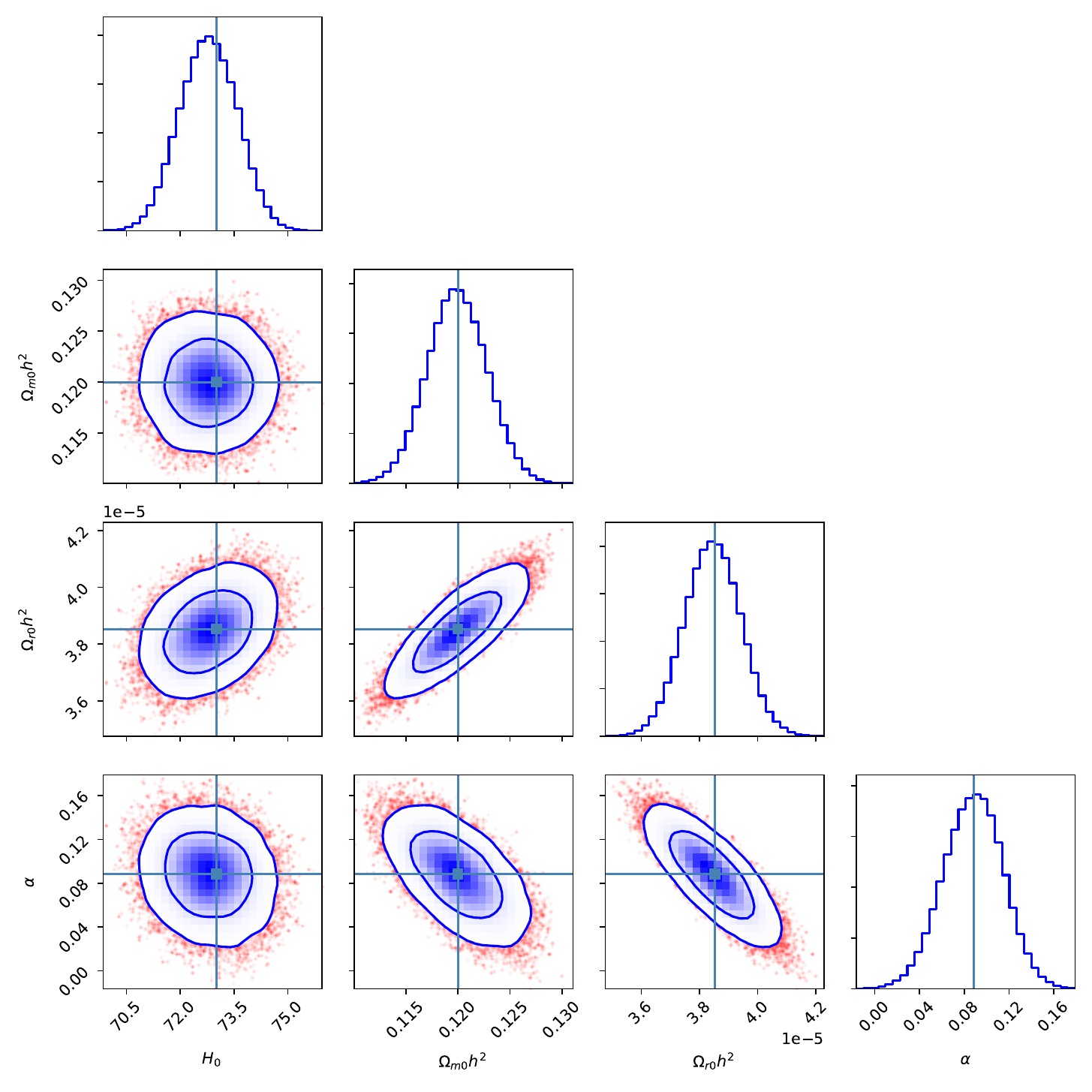}
		\caption{Constraints on the KHDE model parameters from 1$\sigma$ to 2$\sigma $ confidence level. }
		\label{khde}
	\end{figure}

	For the sake of comparison, we have also conducted an optimization process for the wCDM model, which features a constant equation of state parameter $w$, and has an equivalent number of parameters as the KHDE model. The fitting results for the $\Lambda$CDM model are also included in Table \ref{fit}. For a comprehensive account of the fitting methodology, please refer to Refs.\cite{Yang2022,Yang2023}. Figure \ref{khde} illustrates the constraints on the parameters of the KHDE model at the 2$\sigma$ confidence level.
	\begin{table}[!h]
		\centering
		\resizebox{!}{!}{
			\renewcommand{\arraystretch}{1.5}
			\begin{tabular}{cccc}
				\hline
				\hline
				Parameter            & KHDE    & wCDM           & $\Lambda$CDM                        \\
				\hline
				$\Omega_{m0}h^2$      & $0.120^{+0.003}_{ -0.003}$      & $0.115^{+0.003}_{-0.003}$  & $0.106^{+0.002}_{-0.002}$\\
				$\Omega_{r0}h^2/10^{-5}$  & $3.85^{+0.092}_{ -0.092}$           & $4.44^{+0.055}_{-0.055}$ & $4.60^{+0.054}_{-0.054}$   \\
				$H_{0}$               & $72.8^{+0.784}_{ -0.780}$ & $71.4^{+0.719}_{-0.719}$  &  $67.5^{+0.532}_{-0.532}$  \\
				$\alpha$          & $0.088^{+0.025}_{-0.027}$  & -	& -\\
				$w$  				& - &$-1.31 ^{+0.039}_{-0.039}$ & - \\
				\hline
				$\chi^{2}_{min}/dof.$   & $1102.9/1104$   & $1096.3/1104$  & $1172.4/1105$                    \\
				\hline
		\end{tabular}}
		\caption{The best fitting values of the model parameters and their corresponding $1\sigma$ confidence intervals for the KHDE, wLCDM, and $\Lambda$CDM models. Here $h$ is defined as $H_0=100h$ km/s/Kpc. }
		\label{fit}
	\end{table}
	The constraints on the parameter $\beta$ can be obtained from Equ.\eqref{eq:beta} and the fitting results of $\Omega_{m0}$, $\Omega_{r0}$ and $\alpha$ as follows:
	\begin{eqnarray}
		\beta = 0.686^{+0.026}_{-0.027} \,.
	\end{eqnarray}
	The fitting results show that the parameter $\alpha$ in the KHDE model is non-zero at least within the 2-$\sigma$ confidence level, while the parameter $\beta$ is significantly non-zero. Additionally, we find that the fitted value of $H_0 = 72.8$~km/s/Mpc in this model, which is relatively large indicating the absence of the Hubble tension in this model. In contrast, the $H_0$ values in the wCDM and $\Lambda$CDM models are smaller, exhibiting varying degrees of the Hubble tension.
	
\begin{figure}[htbp]
		\centering
		\includegraphics[width=0.6\linewidth]{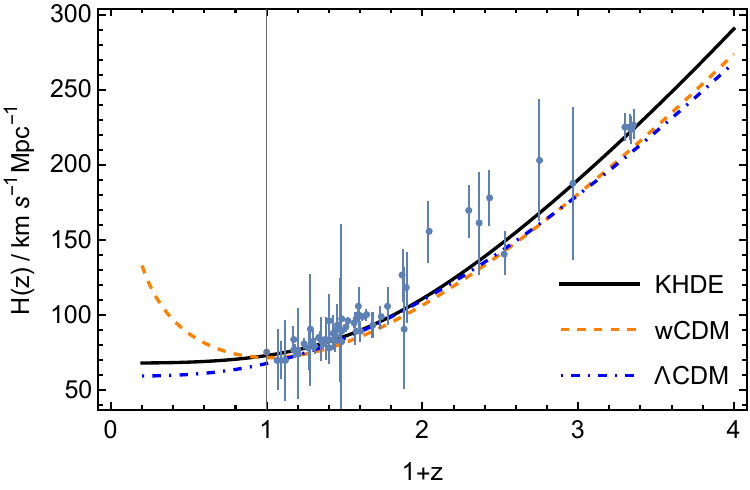}
		\caption{The Hubble parameter's variation with redshift is depicted, with the black solid line, yellow dashed line, and blue dot-dashed line representing the KHDE, wCDM, and $\Lambda$CDM models, respectively. The parameters in the models are set to their best-fit values. The error-weighted variances $\chi^2_{H}$ for the KHDE, wCDM, and $\Lambda$CDM models are $81.65$, $234.75$, and $160.10$, respectively}
		\label{hubble}
	\end{figure}

	As can be seen from Figure \ref{hubble}, the evolution behavior of the Hubble parameter in the three models is somewhat different, especially at recent and future times. The Hubble parameter in the KHDE model is slightly smaller than that in the $\Lambda$CDM model in the early Universe, but at recent times ($z=0$), it is larger than in the other two models. In the future, it will tend towards a constant, which is also the case in the $\Lambda$CDM model. However, for the wCDM model, due to its $w < -1$, the Hubble parameter will continue to increase. From Figure \ref{hubble}, the error-weighted variances $\chi^2_{H}$ for the KHDE, wCDM, and $\Lambda$CDM models are $81.65$, $234.75$, and $160.10$, respectively. It should be noted that the results in Figure \ref{hubble} slightly differ from those in Ref.\cite{Sharov:2018yvz} because Figure \ref{hubble} shows the outcome of joint optimization with all data, whereas Figure 1 in Ref.\cite{Sharov:2018yvz} displays results using only the Hubble data, without presenting the joint optimization results.
	
\begin{figure}[htbp]
		\centering
		\includegraphics[width=0.4\linewidth]{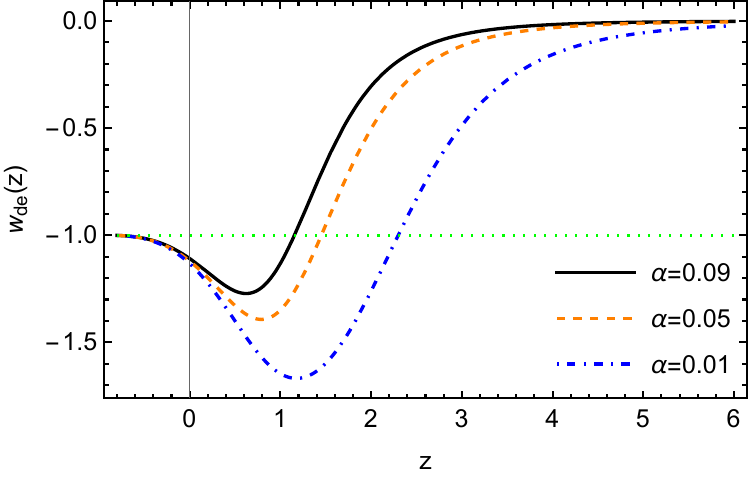}
		\includegraphics[width=0.4\linewidth]{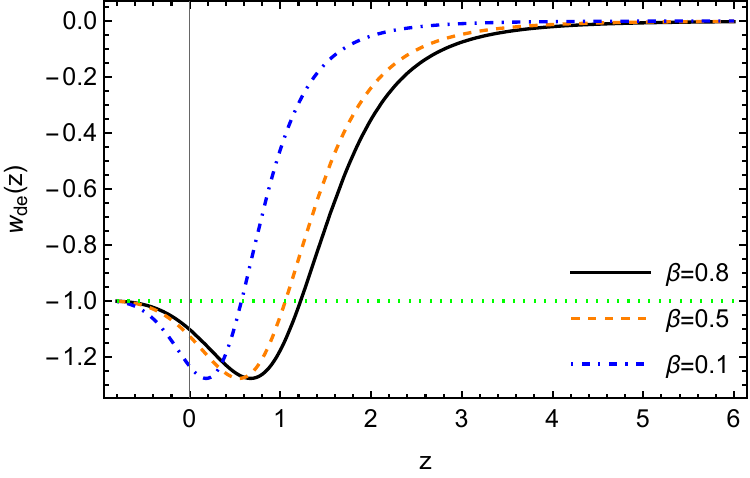}
		\caption{Evolutions of the state equation parameter  with redshift in the KHDE model. Left: $\beta = 0.686$, Right: $\alpha = 0.088$, and all other parameters are set to their best-fit values.
		}
		\label{eos}
	\end{figure}

	In the KHDE model, the state equation parameter of dark energy evolves from $w = 0$ in the early Universe to $w < -1$, and then further to $w = -1$, and the its current value is $w=-1.109$. In Figure \ref{eos}, we compare the evolution under different parameter values. 
	From Equ.~\eqref{eq:rho} , we can also obtain the redshift $z_{EQ}$ at which the energy density of dark matter equals that of dark energy.
	\begin{eqnarray}
		z_{EQ} =\left[ \frac{\beta }{2(1-2\alpha) \Omega_{m0}^2} \right]^{1/6}-1 \approx 0.419 \,,
	\end{eqnarray}
	with the best-fit values incorporated in the final step of the calculation.

	\section{Solution of the Friedmann equation and the age of the Universe}
	
	From the Friedmann equation Equ.~\eqref{eq:sol2}, we obtain
	\begin{eqnarray}
		a^{1/2}\left[ 1 + \sqrt{1 +4\beta  (1-\alpha)\Omega_{m0}^{-2}a^{6}} \right]^{-1/2}  da =  \sqrt{\frac{\Omega_{m0}}{2(1-\alpha)}}H_0dt \,.
	\end{eqnarray}
	Define
	\begin{eqnarray}
		x= 2\sqrt{\beta (1-\alpha)} \Omega_{m0}^{-1}a^{3}\,,
	\end{eqnarray}
	then we get
	\begin{eqnarray}
		x^{-1/2} \left[ 1 + \sqrt{1 +x^2} \right]^{-1/2}  dx= 3 \left(\frac{\beta}{1-\alpha}\right)^{1/4}H_0dt \,.
	\end{eqnarray}
	After integrating on both sides of the above equation, the age of the Universe at time $a$ is given by
	\begin{eqnarray}
		t(a) = \frac{2}{3H_0} \left(\frac{\beta}{1-\alpha}\right)^{-1/4}  \bigg(  -y + \tan ^{-1}y
		+ \tanh ^{-1}y \bigg)   \,,
	\end{eqnarray}
	where
	\begin{eqnarray}
		y 
		&=& \sqrt{\frac{1-\sqrt{x^2+1}+x}{1+\sqrt{x^2+1}-x}} \,.
	\end{eqnarray}
	Therefore, it yields an estimated age of the Universe of $t_0 = 14.2$~Gyr based on the best-fit parameters, which satisfies the constraint imposed by the stellar age bound: $t_0 > 11 \sim 12$ Gyr. In other words, the age of the Universe should be greater than that of any other celestial bodies within it.

	\section{Prediction on the number of strong gravitational lenses}
	Galaxy-galaxy strong gravitational lensing is a powerful tool for constraining astrophysical and cosmological parameters due to its sensitivity to the properties of galaxies as well as the geometry of the Universe. A different dark energy model may predict different expansion history of the Universe and affect the spacial distribution of galaxies along the line-of-sight, thus changing the lensing probability. 
	
	\begin{figure*}[!htbp]
		\centering
		\includegraphics[width=0.7\textwidth]{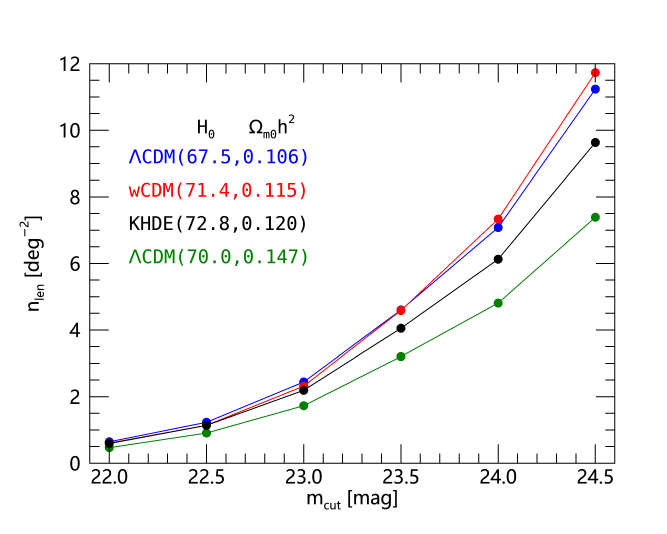}\\
		\caption{The number of galaxy-scale lenses per square degree for Euclid Wdie VIS survey as a function of magnitude cut, where the SNR is larger than 20, the lensing magnification is larger than 5, and no lens light is considered.}\label{Fig:lens}
	\end{figure*}
	
	The on-going Euclid Wide Survey is expected to be able to identify $\mathcal{O}(10^5)$ galaxy-scale lenses, which may help us distinguish between different dark energy models based on the lensing statistics. As a test, we use the analytical model for galaxy-galaxy strong gravitational lensing statistics proposed by \cite{Ferrami:2024obm} to calculate the number density of detectable lenses in KHDE, wCDM and $\Lambda$CDM models with the best-fit parameter values from Section III. In the calculations, we adopt the velocity dispersion function by \cite{Mason_2015} and the UV luminosity function by \cite{Bouwens_2022}. The minimum signal-to-noise ratio and magnification of bright image are set to be 20 and 3, respectively. We show in Figure \ref{Fig:lens} the number density of identifiable galaxy-scale lenses as a function of magnitude cut for the three dark energy models investigated in this paper. As expected, the number density of detectable lenses increases with the magnitude cut of the survey. More importantly, the differing rising tendencies illustrate the strong dependence of lensing occurrence rate on the parameters of dark energy models. It demonstrates that lens counts, with different magnitude cut in Euclid survey, can provide powerful constraints on cosmological parameters and serve as a complementary tool to distinguish the KHDE model from others.
	
	\section{Statefinder diagnostic}
\begin{figure}[htbp]
		\centering
		\includegraphics[width=0.3\linewidth]{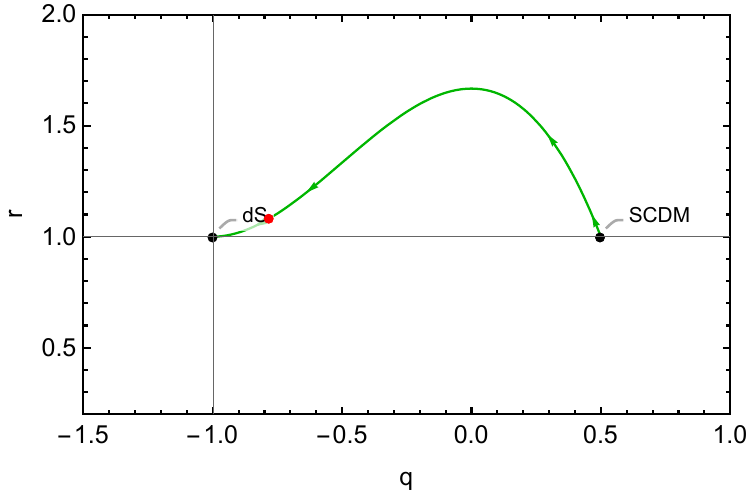}
		\includegraphics[width=0.3\linewidth]{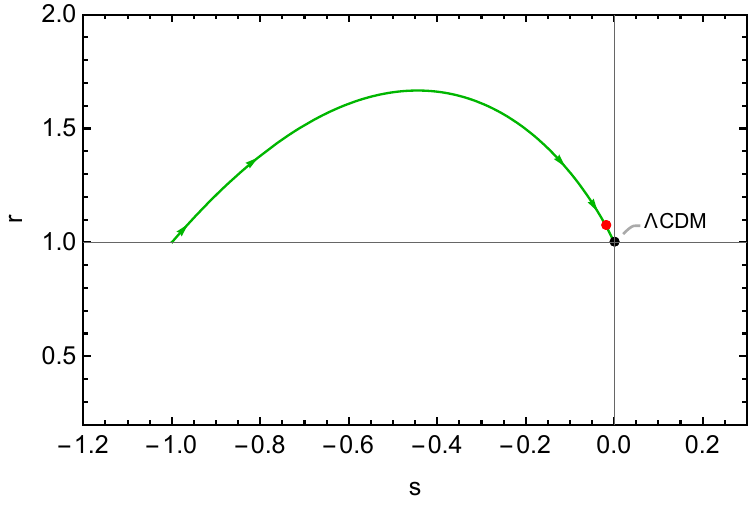}
		\includegraphics[width=0.3\linewidth]{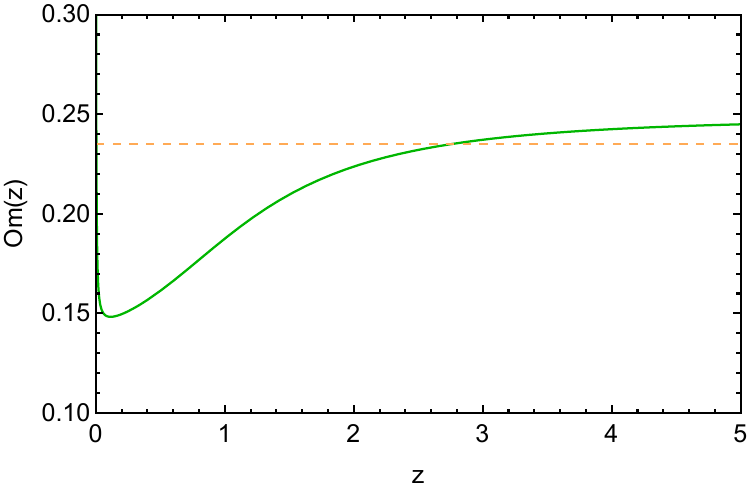}
		\caption{
			Left: The $q-r$ plane, where the black dot at $(-1,1)$ signifies the de Sitter Universe, and the point $(0.5,1)$ denotes the standard cold matter (SCDM) Universe. Middle panel: The $r-s$ plane, with the black dot at $(0, 1)$ indicating the $\Lambda$CDM model The solid dots along the lines representing the current state of the KHDE model. 
			Right: The $Om$ diagnostic for the KHDE model, where the red dashed line indicates the $Om$ diagnostic for the $\Lambda$CDM model.
		}
		\label{dia}
	\end{figure}

	The statefinder diagnostic is utilized through the parameters $\{q, r, s\}$, which are deduced from the higher-order derivatives of the scale factor in the context of various dark energy models \cite{Feng:2008rs}. These parameters are defined as follows:
	\begin{equation}
		q\equiv-\frac{\ddot a}{aH^2},\quad r\equiv\frac{\dddot{a}}{aH^3},\quad s\equiv\frac{r-1}{3(q-1/2)}\,.
	\end{equation}
	For the KHDE model, these parameters are given by 
	\begin{eqnarray}\label{eq:dea}
		q = -1- \frac{\dot H}{H^2} =-1 + \frac{3}{2} \Omega_{m0} a^{-3}\bigg[ \Omega_{m0}^2a^{-6} +4\beta (1-\alpha)  \bigg]^{-1/2} \,.
	\end{eqnarray}
	\begin{eqnarray}
		r 
		&=& 1+\frac{9}{2}\Omega_{m0}^2 a^{-6}\bigg[ \Omega_{m0}^2a^{-6} +4\beta (1-\alpha)  \bigg]^{-1} \left\{ 1-\Omega_{m0} a^{-3}\bigg[ \Omega_{m0}^2a^{-6} +4\beta (1-\alpha)  \bigg]^{-1/2}\right\} \,,
	\end{eqnarray}
	and 
	\begin{eqnarray}
		s = -\Omega_{m0}^2 a^{-6}\bigg[ \Omega_{m0}^2a^{-6} +4\beta (1-\alpha)  \bigg]^{-1} \,.
	\end{eqnarray}
	Additionally, the $Om$ diagnostic is another useful tool, defined as:
	\begin{eqnarray}
		Om(a) \equiv \frac{E^2(a)-1}{a^{-3}-1} \,.
	\end{eqnarray}
	For the wCDM model, the expression simplifies to:
	\begin{eqnarray}
		Om(a) = \frac{\Omega_{m0}a^{-3}+(1-\Omega_{m0})a^{-3(1+w)}}{a^{-3}-1} = \Omega_{m0} + (1-\Omega_{m0})\frac{a^{-3(1+w)}-1}{a^{-3}-1} \,.
	\end{eqnarray}
	In the specific case of the $\Lambda$CDM model, with $w = -1$, $Om(a)$ reduces to a constant, $\Omega_{m0}$.
	For the KHDE model, the $Om$ diagnostic takes the form:
	\begin{eqnarray}
		Om(a) =  \frac{\Omega_{m0}a^{-3}-2(1-\alpha)+  \sqrt{\Omega_{m0}^2a^{-6} +4\beta(1-\alpha)}    }{2(1-\alpha)(a^{-3}-1)}  \,.
	\end{eqnarray}
	The evolutions of these diagnostics are illustrated in Figure~\ref{dia}, from which  it is apparent that there are substantial theoretical differences in the values of these parameters between the KHDE model and the $\Lambda$CDM model. However, ultimately, the KHDE model will behave in a manner similar to the cosmological constant model. Moreover, from Equ.~\eqref{eq:dea}, it can be inferred that the transition from decelerated to accelerated expansion occurred approximately at a redshift of $z \approx 0.843$.

	\section{Conclusions}
	
	In this paper, we introduce a novel dark energy model known as the KHDE model, an acronym for Kaniadakis Holographic Dark Energy. This model draws inspiration from the holographic principle and employs the Kaniadakis entropy as a bound to prevent the Universe from collapsing into a black hole. Departing from convention, we utilize the Hubble horizon, $1/H$, as the infrared cutoff, which successfully accounts for the observed acceleration of the Universe at the current epoch. The KHDE model incorporates just one additional parameter compared to the $\Lambda$CDM model. Our analysis, informed by observational constraints, reveals a Hubble constant of $H_0 = 72.8$ km/s/Mpc, thereby addressing the Hubble tension problem. The transition from dark matter dominance to dark energy dominance occurs at a redshift of around $z \sim 0.419$. Based on the optimal parameter values, we estimate the age of the Universe to be $14.2$ billion years. Additionally, we  predict the number of strong gravitational lenses and perform statefinder and $Om$ diagnostic analyses to further validate and delineate the model's characteristics.
	
	The KHDE model possesses several distinctive features: 1) it does not suffer from the Hubble tension problem; 2) it has only one more parameter than $\Lambda$CDM model; 3) the equation of state parameter  $w$ crosses -1; and 4) the Universe ultimately evolves into a cosmological constant model, thus avoiding the 'big rip' problem. In fact, although our starting point is the approximate Kaniadakis entropy and the holographic principle, we can still ask whether there exists a fundamental theory of gravity in which the Friedmann equations given in Equ.~\eqref{eq:sol2} are derived. We leave this question for subsequent research.

	\section{Acknowledgments}
	CJF  acknowledges the support from NSFC grants No.11105091. WD acknowledges the support from NSFC grants No. 11890691.
	

	\bibliographystyle{unsrt}
	\bibliography{refs}
	
\end{document}